# ARISE: A Granular Matter Experiment on the International Space Station

Tobias Steinpilz,[1, a)] Grzegorz Musiolik,[1] Maximilian Kruss,[1] Felix Jungmann,[1] Tunahan Demirci,[1] Manfred Aderholz,[1] Jonathan E. Kollmer,[1] Jens Teiser,[1] Tetyana Bila,[1] Evelyn Guay,[2] and Gerhard Wurm[1]
[1])*University of Duisburg-Essen, Faculty of Physics, Lotharstr. 1-21, 47057 Duisburg, Germany*
[2])*Baylor University, Department of Physics, One Bear Place 97316, TX 76798-7316 Waco, United States of America*

We developed an experiment to study different aspects of granular matter under microgravity. The 1.5 U small experiment was carried out on the International Space Station. About 3500 almost identical spherical glass particles with 856 µm diameter were placed in a container of 50 by 50 mm cross section. Adjusting the height between 5 and 50 mm, the filling factor can be varied. The sample was vibrated with different frequencies and amplitudes. The majority of the data are video images of the particles' motion. Here, we give a first overview of the general setup and a first qualitative account of different phenomena observed in about 700 experiment runs. These phenomena include collisional cooling, collective motion via gas-cluster coupling, and the influence of electrostatic forces on particle-particle interactions.

## I. INTRODUCTION

In 2018 the experiment 'ARISE (Planet Formation Due to Charge Induced Clustering on the ISS)' was carried out inside the NanoRacks frame in the Kibo module of the International Space Station. ARISE was part of the *Überflieger* program of the German Space Administration[1,2]. The experiments of this program were to coincide with the presence of the European Space Agency astronaut Alexander Gerst during his "Horizons" mission on the ISS. This resulted in a short time span of one year for development and assembly of a 1.5 U experiment within a budget of 15 k€.

The experiment set out to study a phase during planet formation that cannot be explained easily yet. Early on during planet formation mm-particles form by hit-and-stick collisions between smaller grains. However, once that big, particles only bounce off each other in further collisions. This is the so called "bouncing barrier"[3–5]. Millimeter is therefore a critical size for particle evolution in the context of planet formation so details of collisions and potential for further aggregation is important. Brisset et al.[6] e.g. carried out an experiment on the ISS, called NanoRocks, utilising the large time span of microgravity to study the aggregation of different samples at the mm size scale at very low collision velocities.

While limits might be pushed somewhat by choice of material, adhesion forces might not suffice to promote growth of larger aggregates, eventually. However, informal ISS experiments by Love, Pettit, and Messenger[7] showed that larger aggregates can form. They argue that static charge might allow aggregation to proceed. Along the same direction, drop tower experiments by Jungmann et al.[8] showed that even a nearly monodisperse sample of identical glass-beads can acquire a significant static charge which shifts the velocity below which particles stick together by orders of magnitude. The idea of collisional charging and subsequent aggregation led to the ARISE design to study the aggregation and cluster properties in long time microgravity.

Originally intended to study the influence of static charge on aggregation, the experiment turned out to be a versatile instrument to study the physics of a cloud of sub-mm glass spheres under different conditions. In the following we therefore give a first detailed review of the ARISE experiment and then highlight a few scientific phenomena which are included in detail in the different data sets that were generated on the ISS.

## II. THE ARISE EXPERIMENT

### A. Mechanical Assembly

The ARISE experiment is dimensioned to fit in a 1.5 U NanoRacks module measuring 100 x 100 x 150 mm while the overall weight of the experiment is about 1.6 kg including the container. As shown in fig. 1, the mechanical components are mounted on a base plate to simplify the integration in the container. The central part of the experiment consists of a closed sample chamber with an inner volume of 50 x 50 x 50 mm.

It is filled with approximately 3500 nearly identical glass beads (Whitehouse Scientific® MN0881), which is roughly a mono-layer of spheres on the ground. They have diameters of 856 µm with a standard deviation of 15 µm and masses of 858 µg with a standard deviation of 28 µg (single beads measured with a scale) each. See fig. 2 for a size distribution measured directly via microscopy. The narrow deviation of the particles' mean size allows to derive the approximate 3D position of a sphere inside the chamber from the camera images by comparing the measured diameter to the known diameter with around ±5 mm accuracy. The glass spheres can move and interact freely with each other inside the sample chamber.

The height of the observable volume can be varied between 5 and 50 mm by moving the lid via a tooth belt system. For this purpose, a step motor (Nanotec® ST2018) equipped with a 1:25 planetary gear (Nanotec® GPLL22-25) drives two vertical tooth belts coupled by a shaft. Two thin steel sheets link the tooth belts to the lid so the lid can be set in motion at a velocity of typically 5 mm/s. Since the amount of beads is

---

[a)]Electronic mail: tobias.steinpilz@uni-due.de

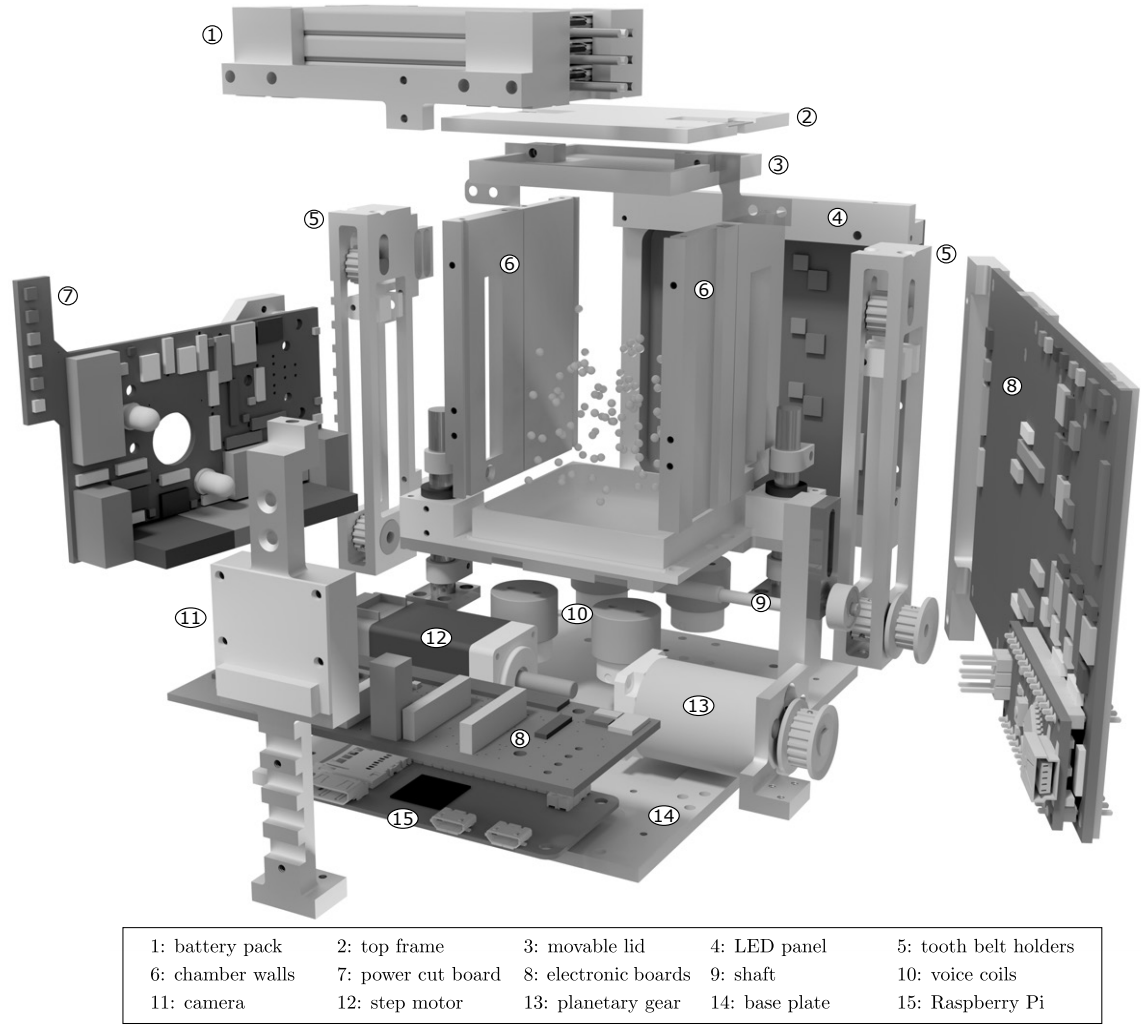

| 1: battery pack | 2: top frame | 3: movable lid | 4: LED panel | 5: tooth belt holders |
| 6: chamber walls | 7: power cut board | 8: electronic boards | 9: shaft | 10: voice coils |
| 11: camera | 12: step motor | 13: planetary gear | 14: base plate | 15: Raspberry Pi |

FIG. 1. Exploded view of the main parts of the experiment. The power-cut-board (7) is mounted on the camera (11).

constant, this effectively leads to a variation of the volume filling factor $\Phi = 46.5 \cdot h^{-1}$ [%], where $h$ is the height of the lid in mm above the bottom of the chamber. This allows tuning the volume filling between around 1-9 %, which is well below random loose packing for a solid[9].

The whole sample chamber can be shaken by 4 voice coils (PBA Systems® CVC16-SF-5) mounted underneath. Typical shaking frequencies are 0.25 - 50 Hz, whereas the maximum stroke is up to 5 mm but varies depending on the frequency, see fig. 3 for more details. The vibrations induce collisions among the glass beads. The number of collisions can be increased by moving the lid to the lowest position and thus increasing the volume filling.

To check for potential charged spheres, it is possible to apply a voltage of 3.3 V to the side walls of the sample chamber. These act like a capacitor allowing to accelerate charged particles. While the chamber walls are milled from aluminum just like most of the other components of the setup, the bottom part as well as the top frame of the chamber are made of Polyoxymethylen to ensure electrical insulation. Although the resulting acceleration on an averagely charged grain is on the order of $10^{-6}$ g and therefore two magnitudes below the g-jitter the low voltage was chosen because of safety restrictions. Optimally a much higher voltage should be applied to the chamber walls. Therefore further analysis is needed to separate the resulting acceleration from the g-jitter noise.

The lid and a funnel-shaped inset on the bottom of the sample chamber are coated with the sample beads to ensure only mutual collisions of the particles during agitation.

Further components shown in fig. 1 are a Raspberry Pi Camera V2 module with a 8 Mpx sensor, which observes the interaction of the glass beads, and a 3x3 matrix white LED panel providing the light source. The given geometry results in a resolution of $f(z) = 1331.67 \cdot (0.0612227 + z)^{-1}$ [px/m], where $z$ is the depth inside the sample chamber in meters. A lithium polymer battery pack (LiPo) ensures sufficient power supply. The whole experiment is controlled and monitored by a Raspberry Pi Zero V1.3 (Pi) and proprietary printed circuits boards (PCB).

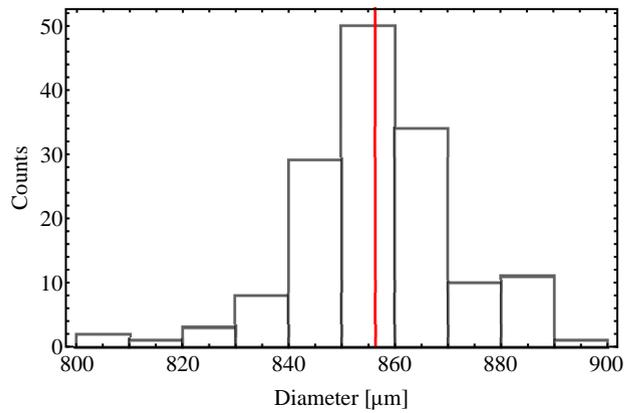

FIG. 2. Size distribution of the glass spheres used, measured by an optical microscope. The red line represents the mean diameter of $856\,\mu$m.

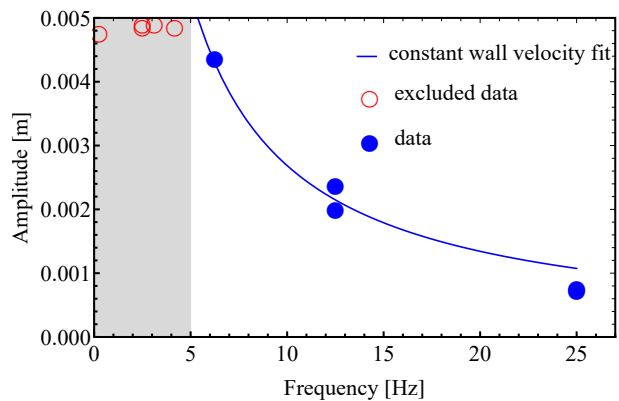

FIG. 3. Measured vibration amplitude $A$ of the chamber over frequency $f$. The blue line is a fit with $A = 0.0268 \cdot f^{-1}$ [m] which corresponds to a constant wall velocity. The red dots are excluded from the fit since the amplitude is limited by the maximum stroke. The agitation amplitude for higher frequency could not be measured within the limits of our spatial and temporal resolution. The sample's absorbed kinetic energy can directly be measured from the particle movement.

### B. Electronics

The NanoRacks platform provides the power supply and data transfer via a single USB3.0 type-B connector. It allows a maximal drain of 900 mA at 5 V and the data-handling via a mass storage device protocol or the use of a FTDI®-serial-chip. Since the Pi has a USB On-The-Go (OTG) port, it is possible to let it act as a USB-slave to emulate a mass storage device. This makes the data handling between the experiment and the NanoRacks frame easy by simply connecting the data wires of the OTG port to the corresponding USB3 connector pins.

To meet the maximum current requirements and safety issues we developed a modular power grid split on two PCBs, which typically supply all loads autonomously. The whole electronics uses a common ground (GND). Both 90 $\mu$m copper layers of the PCBs are also GND giving additional electromagnetic interference protection. There are a few main loads, the Pi, the voice coils, the stepmotor for lid-movements and the back-light-LEDs, which may consume power simultaneously and therefore exceed the maximum peak current of 900 mA. As a result we chose to split them in sub-grids and give each a LiPo (VARTA® LPP 503759 8H) with 1400 mAh to buffer these times of peak power consumption. Each LiPo has its own charger supplied via the USB3 connector. All chargers utilize the MCP73831 chip and supply a charge current of 100 mA except the LiPo, which acts as a uninterruptible power supply for the Pi which is charged with 500 mA to fit to the power consumption of the Pi. As a step-up driver we used the TPS61090 to create stable 5 V input for the Pi out of the 3.7 V LiPo output. A poly-fuse and a relay are placed directly at each LiPo as safety measure - these are located on the power-cut-board which is mounted on the camera (compare fig. 1).

Also mounted on the power-cut-board and pointing at the sample chamber, two 375 nm UV-LEDs and a small fan are powered directly via the ISS-power-supply and are switchable with an n-channel MOSFET.

To drive the stepper motor or the voice coils we used the low-voltage driver DRV8834 for each. The back-light-LEDs are driven by a SN3218 chip allowing simple adjustments of the intensity via I2C-protocol.

For logging and housekeeping data we added a real-time-clock (RTC - DS3231) and an atmospheric sensor (BME280 - measuring ambient pressure, temperature and humidity).

An Arduino® Nano acts as a watchdog, ensuring a supplied and responding Pi if ISS-power is present. It also measures the LiPo-voltages and switches the loads on request. The communication uses the I2C-protocol with the Pi as master.

Further probing mechanisms like a higher capacitor voltage (better charge measurements), a stereoscopic camera (better depth resolution) or a microphone (as impact sensor) could not be implemented due to space, time, power and budget restrictions.

### C. ARISE in operation

During the operational time, ARISE was connected to the NanoRacks frame in the Kibo module on the ISS. Once per day, an update containing new software files with the experimental instructions for the next 24+ hours was provided to the NanoRacks engineers. The data generated by ARISE were then transferred to a server one day later. The emulation of a mass storage device requires a power cycle of the Pi, since read and write actions must not be performed by two devices (Pi and NanoRacks frame) at the same time. Therefore we have two partitions which are only read- and writable by one device at a time and are switched during start up - this results in an asynchronous execution of commands. Logging of the housekeeping-data and reaction on events or commands (like the drop of LiPo-voltages or a reboot request) are executed each minute by a cron job. The scripting of experiments is handled via a second independent task.

The g-jitter inside the experiment is on the order of $10^{-4}$ g and not isotropic. On a timescale of approximately 5 min, the sample noticeably moves towards the front window and the bottom. This motion is overlapped by smaller isotropic accelerations below $10^{-4}$ g. After half of the operational time, the experiment was displaced to the other side of the NanoRacks frame resulting in a g-jitter pointing towards the back window and the bottom. Due to the RTC it is possible to time experiments fitting to the crew-schedule in times where the isotropic g-jitter is minimal – usually between 1 a.m. and 5 a.m. UTC. Furthermore it is possible to use the "Principal Investigator Microgravity Services" to obtain g-jitter data and link this to our measurements.

In total about 700 individual experiments were carried out during which we recorded around 1500 highly compressed .h264 1080p-videos, resulting in roughly 200 GB of data.

Around 350 experiment runs were dedicated to study collisional charging and how it alters the aggregation process. In these experiments we captured 1080p-videos of the vibration and expansion phase with 30 frames per second (fps) and took time-lapse videos with 2 fps - which is the lowest value the cam can capture continuously in the video mode. For the vibration phase (duration's between 5 and 90 minutes, varying frequencies) the volume was reduced around the minimum chamber volume - this increases mutual collisions and reduces wall interaction of the sample. Afterwards we usually expanded the volume to the maximum or two thirds. Occasionally we applied the electrical field via the chamber walls, used the UV-LEDs to alter the charge or re-agitated the sample.

In the second month the other 350 experiment runs were focused on experiments on granular dynamics. Here we captured videos with a broader variety of fps – 2 to 30 fps 1080p-videos and 90 to 120 fps in 640p-videos which is the cropped center in the middle of the chamber. The chamber volume and agitation frequency varied over the full range, where the vibration duration spans from 1 minute, to study the cooling afterwards, to 2 hours, for "time-lapse" studies.

Of course both sets of data will show features that are of interest for either area of research.

## III. FIELDS OF STUDY

This paper is focused on the review of the instrument as detailed above. Data analysis has only begun and is ongoing. However, we will outline a few phenomena that are immediately visible in the data. We note though that this is only meant to show directions as data on each of these topics, being part of the present data set, will require in depth analysis far beyond this short listing.

### A. Charge and Adhesion

*Charging, motion in E-fields and charged aggregation:* The basic motivation of ARISE was to study collisional charging and its effects on particle collisions and cluster formation. Jungmann et al.[8] studied impacts of charged grains onto metal targets in strong electric fields. Our experiments can significantly extend these studies. Fig. 4 shows a sketch of the experiment chamber with elements important for electric fields and charges. In general the chamber might be divided into two regions. In the inner (blue) region the g-jitter of the space station dominates the particle motion. Close to the walls (red region) particle motion differs significantly from simple g-jitter motion. Superimposed insets on the top and left e.g. show examples of trajectories of grains being attracted by the walls. As these walls are metal (left) as well as isolating (top), different motions can be studied. The motion depends on the electrical field. A small electrical field was applied between left and right wall. Through the top and bottom, which are not metal also external fields e.g. generated from the experiment electronics can enter. The field close to the center of the metal walls is supposed to be homogeneous, while the fields especially at the corners at the top and bottom are likely inhomogeneous. This generates different particle motions, as inhomogeneous fields also attract dipoles, while homogeneous fields only align them. In addition, mirror charges on the electrodes provide attraction. This is a complex mixture of electrical fields and charge distributions not known yet and a consistent picture has to be set from a sufficient number of trajectories analyzed.

The charges also change the way particles collide among themselves and how they grow to clusters. An example of such collisions and the formation of a small cluster is also shown in fig. 4 (center).

*Collision properties of large grains:* In more detail, collision properties of large grains in mutual collisions - grain onto grain - can be studied. Studying pairwise collisions allows testing extremes of given sticking models e.g. by Thornton and Ning[10] or Musiolik et al.[11,12]. It might also allow to compare sticking velocities of charged and uncharged grains. The results can be applied to collisional growth in planet formation at the bouncing barrier[3–5]. Depending on the increase in sticking velocity, aggregates can proceed to grow larger.

Estimates for sticking velocities for uncharged grains are on the order of 1 mm/s for central collisions assuming a surface energy of $\gamma = 0.3 J/m^2$ for the glass particles[8]. If the effective surface energy is much lower as e.g. implied by work by Demirci et al.[13] then the sticking velocity would also be lower. This cannot readily be studied on the ground though. In any case, Jungmann et al.[8] found that charges shift the sticking velocity of glass spheres with a metal wall due to mirror charges. They observed sticking velocities of grains half the size as used here being in the cm/s range. They showed that the sticking velocity depends on the grain charge.

A first analysis of a few tens of collisions in ARISE showed that also here the sticking velocity is higher than calculated. We found values being somewhat lower than 1 cm/s in three experiment runs partly analyzed - see fig. 5. It has to be noted that grain-grain collisions are different from the collisions studied by Jungmann et al.[8] as there is no mirror charge. A more detailed analysis of this process is needed. A similar situation to the mirror charge effects is found though, if two grains with opposite charges approach each other.

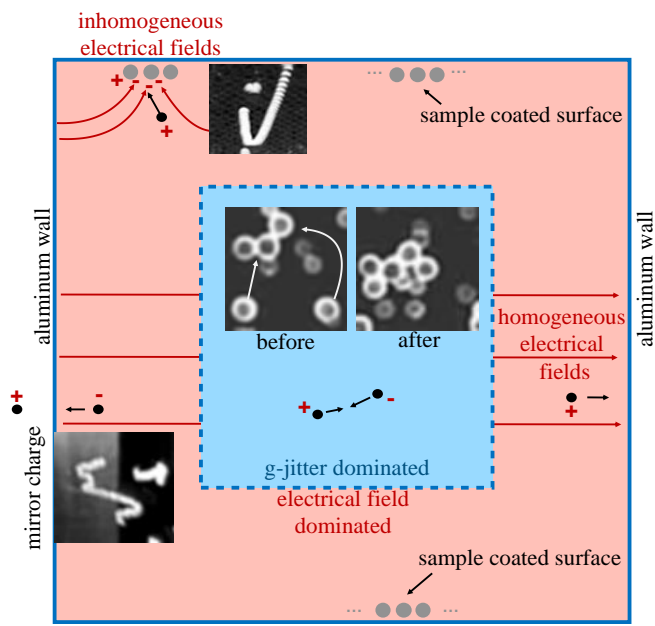

FIG. 4. Sketch of the experiment chamber and different local particle motions of charged grains with real data insets of particle superpositions visualizing trajectories. Red: region close the walls, particle motion influenced by external electrical fields. Blue: inner region, particle motion dominated by g-jitter and electrical fields between particles close by. Examples: (top left) particle trajectory influenced by a inhomogeneous electricla fields on the particle coated insulator surface. (bottom left) trajectory of grain in homogeneous external field and field of induced mirror charge on the aluminum wall. (center) formation of a small cluster by collisions with charged grains, e.g. visible in detail (not shown) as curved or accelerated approach to each other.

Due to the complexity it is not clear yet what the absolute charges on the grains are. In this context it also has to be noted that the radiation conditions in ARISE are different from ground conditions as e.g. the cosmic radiation is about 100 times larger on the ISS compared to the ground, which might influence the charge state.

In any case, the high sticking thresholds imply that the grains are charged and that charges promote dimer formation.

*Stability of dimers and aggregates:* As just seen, dimers should be more stable due to charge. Once dimers formed, one might approach their stability from a different perspective. Collisions are rarely central or symmetric. Therefore, after collisions dimers are often rotating more or less rapidly. An example of a rotation is seen in fig. 6. Rotation leads to a well defined centrifugal force $F_r = 8\pi^2 mrf^2$ acting between the two grains (each with the radius $r$ and the mass $m$) of a dimer. The rotation frequencies $f$ are therefore means to estimate the cohesive force between two grains - charged or uncharged. From the example of fig. 6 with $f = 0.69$ Hz we get $F_r = 3 \cdot 10^{-5}$N. Compared to that, the cohesive force of an uncharged grain is $F_c = 3/2\pi\gamma r$ of, with $\gamma = 0.3 J/m^2$, $F_c = 10^{-4}$N. This is stronger than the centrifugal force. So this example would be consistent with normal adhesion. Faster rotators or the frequency of their occurrence harbor information on contacts which are more stable than usual and they will be analyzed in the future.

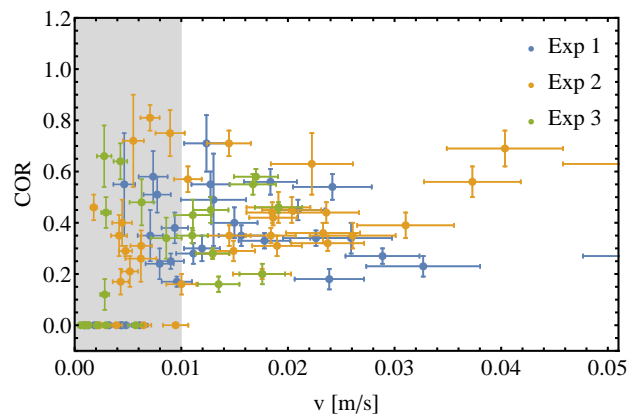

FIG. 5. From the on-orbit data we obtained particle tracks via image analysis. From the collision observed in the tracks we deduce the coefficient of restitution (COR) for potentially charged particle-particle collisions for different initial relative velocities. Note the critical sticking velocity in the range of 1 cm/s (gray region - sticking events are observed). The strong scattering is due to the unknown share of rotational energy and the uncertainty in the z-axis kinetic energy. Also this z-axis uncertainty causes the large error bars for the relative velocity - we assumed this error to be lower than 10 % which corresponds to an uncertainty in z of around ±5 mm. The residual error originates from the fitting uncertainty of the particle tracks.

Not as straightforward, it also occurs frequently that larger aggregates in motion and rotation break up. This also holds information on the sticking properties and resistance to torques as sometimes smaller parts roll over connecting spheres in rigid rotation.

*Collisions of aggregates:* In clusters, grains stick together weakly. Not shown here, we observe aggregates dissolve into smaller fragments and monomers in collisions with other clusters or monomers. A number of different aspects can be studied from this. The velocities and directions of fragments can be measured. This results in an energy balance of impact energy, distributed as kinetic energy of the fragments and dissipated during the collision. This also relates to the aspects mentioned above. It is also a basic process in preplanetary collisions. The data might contribute to the large data base of fragmentation under different conditions relevant for planet formation and small bodies in the solar system (e.g. Brisset et al.[14]).

### B. Granular effects

We start by noting here that we are not the first to study granular media under microgravity aiming for long duration weightlessness. Aumaître et al.[15] for example also report experiments on parabolic flights in the context of the VIP-Gran facility dedicated to be sent to the space station this year (2019).

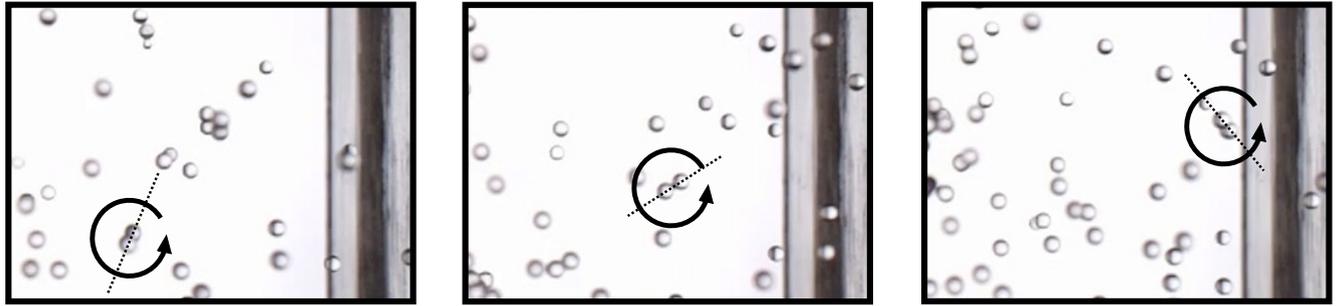

FIG. 6. Example on-orbit image sequence showing the rotation of a dimer at 0.69 Hz ($\Delta t = 0.33$ s between images). From this rotation frequency we can derive a centrifugal force. The obtained data shows hundredths of such dimeres, therefore a lower limit of the sticking force can be determined with further analyses.

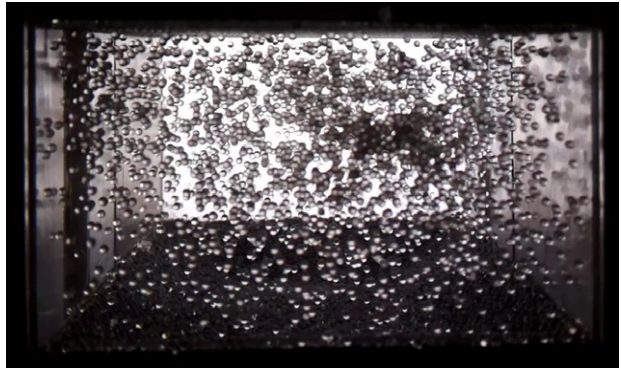

FIG. 7. Dynamic clusters form at specific locations within the chamber, depending on the chamber geometry, filling factor and agitation (here: 50x50x30 mm, $\sim 1.5\,\%$ and $\sim 25$ Hz).

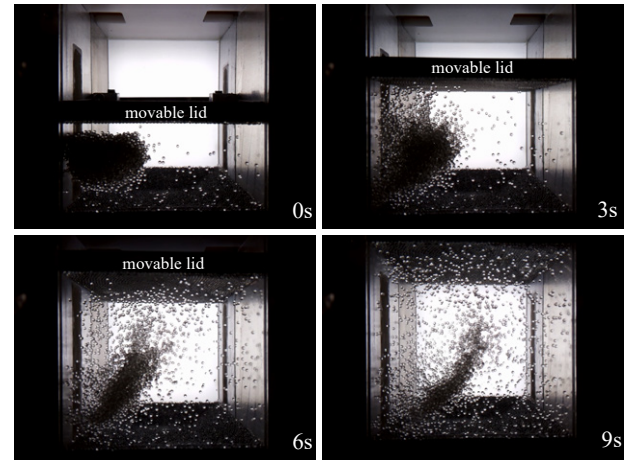

FIG. 8. Image sequence of a giant melting cluster due to the impacts of agitated grains.

As noted above ARISE was not designed to study granular phenomena in a general sense. However, it turned out to show some granular effects listed here. Keep in mind that these effects could not only be granular effects but also influenced by the present charge on the particles.

*Cluster formation and melting:* We observed that agitating the system leads to motions favouring the formation of clusters in certain locations of the chamber. Here, clusters might not necessarily be adhering to each other but just be local dense particle concentrations. Such effects have been studied before by Opsomer, Ludewig, and Vandewalle [16], Noirhomme *et al.* [17]. An example from our experiments is shown in fig. 7. The formation as well as the location of such clusters depends on the agitation frequency and chamber dimensions. The latter changed by moving the top wall. Noirhomme *et al.* [17] give a phase diagram for whether such clusters should occur or the system is a granular gas. This depends on the packing fraction and the system size. Our data extend their data set to larger systems and the example shows that a more detailed analysis of the available data might provide further threshold conditions.

Once a cluster has formed, the wall agitation can be switched on again and the melting of the cluster can be observed in 3-dimensional sample as opposed to previous experiments on 2d-systems [18]. An example of this melting process is shown in fig. 8 where a giant cluster (approximately over 3000 particles) melts up due to the induced energy by fast particles as described in a similar experiment by Katsuragi and Blum [19]. This is a different approach to the one from section "Collisions of aggregates" from above where the effect as well as the analysis may overlap - Our upcoming publication by Musioli *et al.* [20] shows a way to utilize the Shanon Entropy to quantify these processes.

*Granular gas, granular convection and collisional cooling:* A classic granular experiment is to observe how the kinetic energy from vibrating walls is distributed in the system. An equilibrium occurs between energy input and loss due to dissipative collisions [21–23]. In ARISE we can quantify the velocity distribution of the grains as a function of the excitation parameters and the (variable) dimensions of the experiment container and therefore of the volume filling (packing frac-tion). This can either be done by directly tracking the particle motion or by utilizing particle image velocitmetry (PIV) as implemented by e.g. PIVLab[24] to get a flow field. Therefore we can measure the energy input by the walls and the dissipation through the volume directly without deriving these values from the agitation itself.

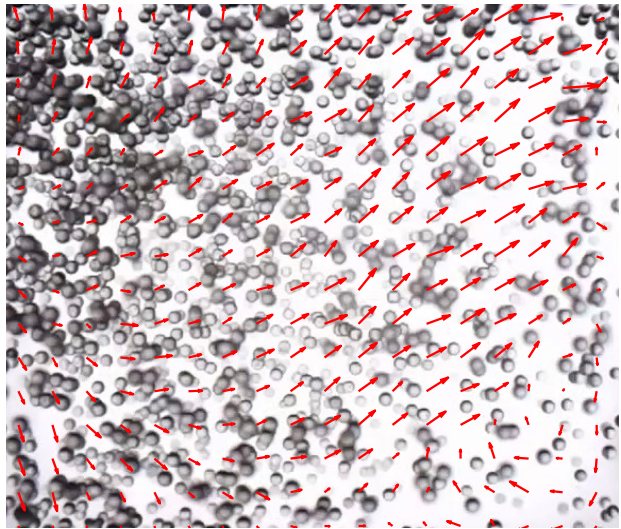

FIG. 9. Example segment inside the chamber of a convection like motion of grains between two vibrating plates. The effect is visualized by superimposing a representation of the flow field inside the chamber (red arrows) calculated using PIV.

A granular gas is continuously losing energy by collisions and cools[25]. This is a main mechanism for structure formation after the system was initially agitated by the vibrating walls. Through repeated collisions grains eventually become so slow that only the g-jitter of the space station keeps them moving. How particle charge affects granular cooling is an open question but significantly changes the cooling and structure formation as e.g. suggested by Singh and Mazza[26]. Here, ARISE is well suited to provide experimental data.

In a vibrated sample with walls shaking differently, collisions with one of the walls can supply more kinetic energy to the grains than collisions with one of the other walls. This can lead to granular convection. An example flow field (calculated via PIV) is shown in fig. 9.

### C. Hydrodynamic effects

It is important to note that the particle sample is surrounded by air at an ambient pressure of about 1000 mbar, therefore also particle-gas interactions are important. Collisionsal cooling of the granular system cannot be separated from cooling by gas drag in our system. Velocities of grains significantly decrease on passage between different walls and between collisions among each other. It will be interesting to see how granular cooling hands over to gas drag and how both influence the motion and cluster formation.

In this context the friction time is important. While it can be calculated for an individual spherical particle, the coupling of a cluster to the gas is not straight forward to be quantified. Depending on local particle density collective effects set in[27]. Especially the remaining g-jitter might come in handy here as it mimics a very slow sedimentation and equilibrium velocities can be measured. So the experiments also offer information here.

Cluster forming in the experiments sometimes come in very systematic patterns. With a slow gas flow induced by the fan cooling the experiment monolayer sheets of grains form reproducible as seen in fig. 10. This effect is tied to gas motion but deeper analysis is required to explain these structures. So this is currently only a curious phenomenon.

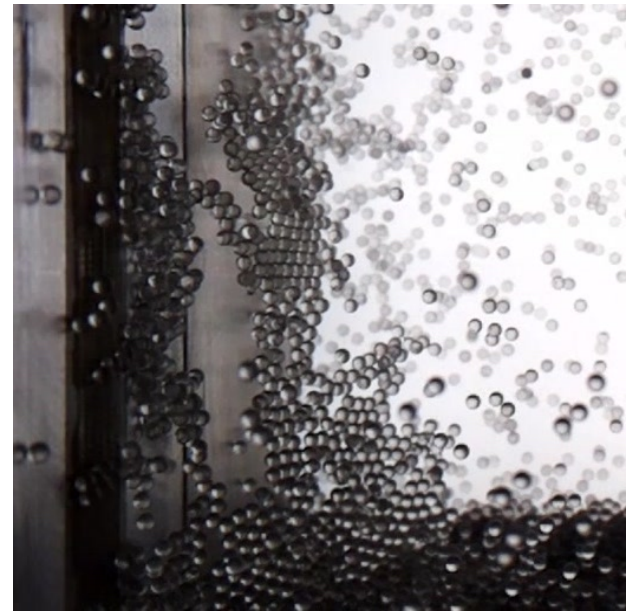

FIG. 10. Monolayer sheets of grains form in some experiments. Note that these sheets are not at the windows but form in the center part of the chamber.

## IV. CONCLUSIONS

We implemented a granular matter experiment for the ISS capable of agitating a sample of glass beads in various ways. As outlined above, considering different initial conditions and boundary conditions, a large number of effects can be observed during the evolution of the experiment.

There probably will be an improved experiment version with less power, data and space constrains in the future. Beside a stronger applicable electrical field on the capacitor walls we would like to improve the optical observation further by adding a second camera.

As the data will be analyzed, this work will serve as a review of the underlying experiment for reference, allowing the wealth of scientific problems to be treated with more focus.

## V. ACKNOWLEDGEMENTS

This work was funded by the DLR Space Administration with funds provided by the Federal Ministry for Economic Affairs and Energy (BMWi) under different grants. Access to the International Space Station was provided under grant 50 JR 1703. Preparatory work was supported by grant 50 WM 1542. F. Jungmann and T. Demirci are supported by grants 50 WM 1762 and 50 WM 1760, respectively. M. Kruss is funded by DFG grant WU 321 / 14-1. We also thank "DreamUp" and "NanoRacks" for their financial and technical support. We kindly thank "Softwareentwicklung Recktor" for providing parts of the software for ARISE. E.Guay also acknowledges funding by the German Academic Exchange Service (DAAD) RISE.

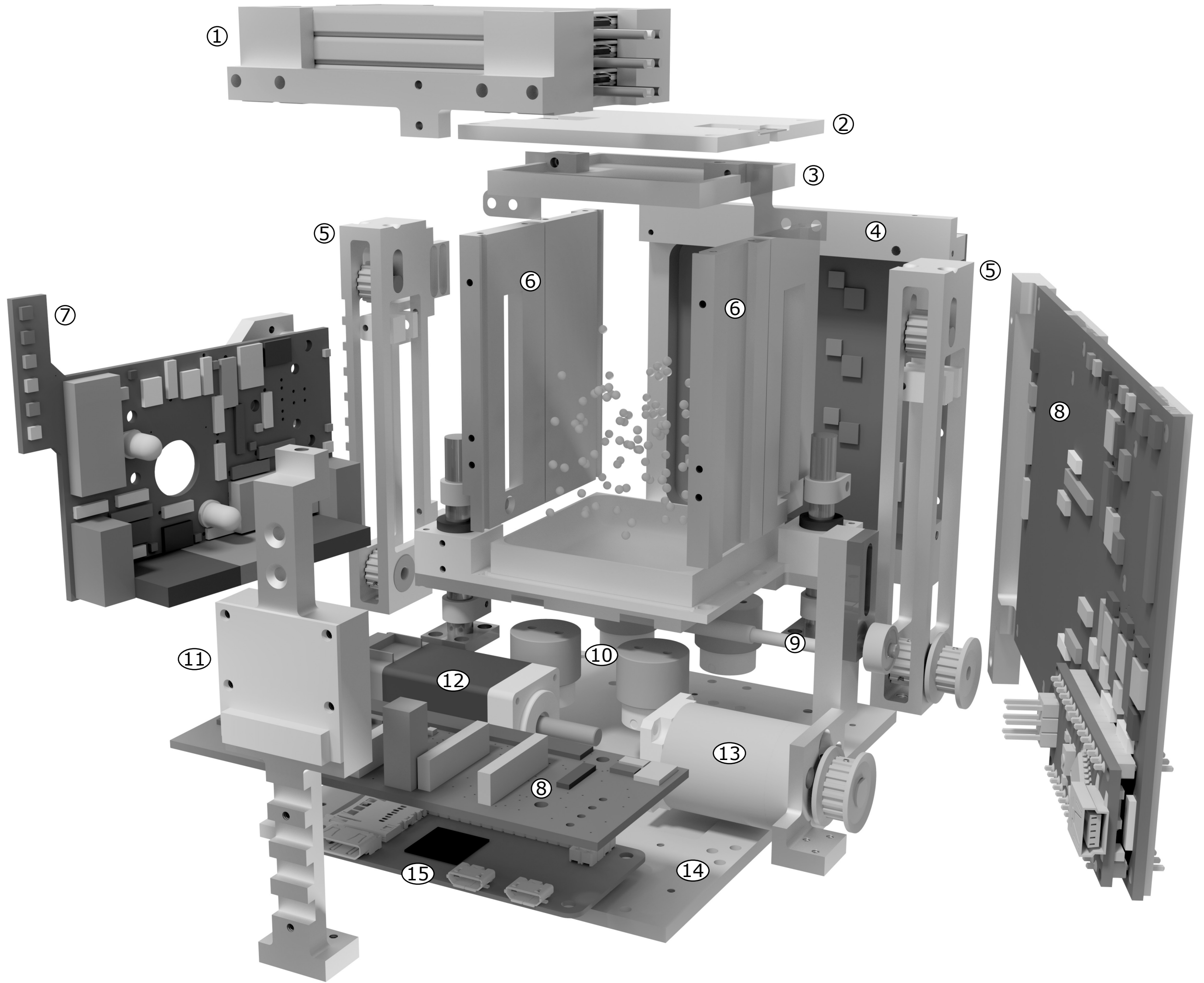

| 1: battery pack | 2: top frame | 3: movable lid | 4: LED panel | 5: tooth belt holders |
| 6: chamber walls | 7: power cut board | 8: electronic boards | 9: shaft | 10: voice coils |
| 11: camera | 12: step motor | 13: planetary gear | 14: base plate | 15: Raspberry Pi |

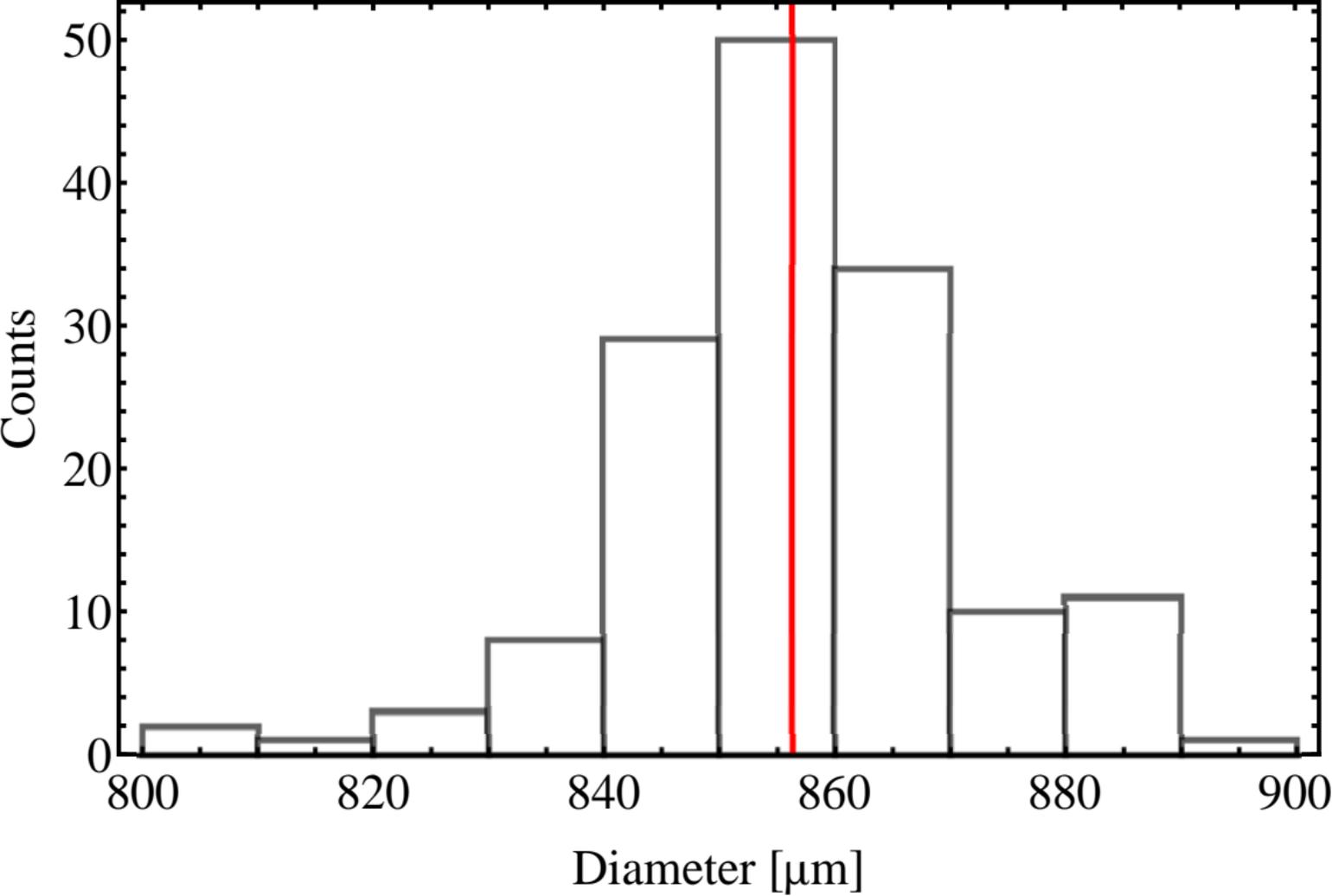

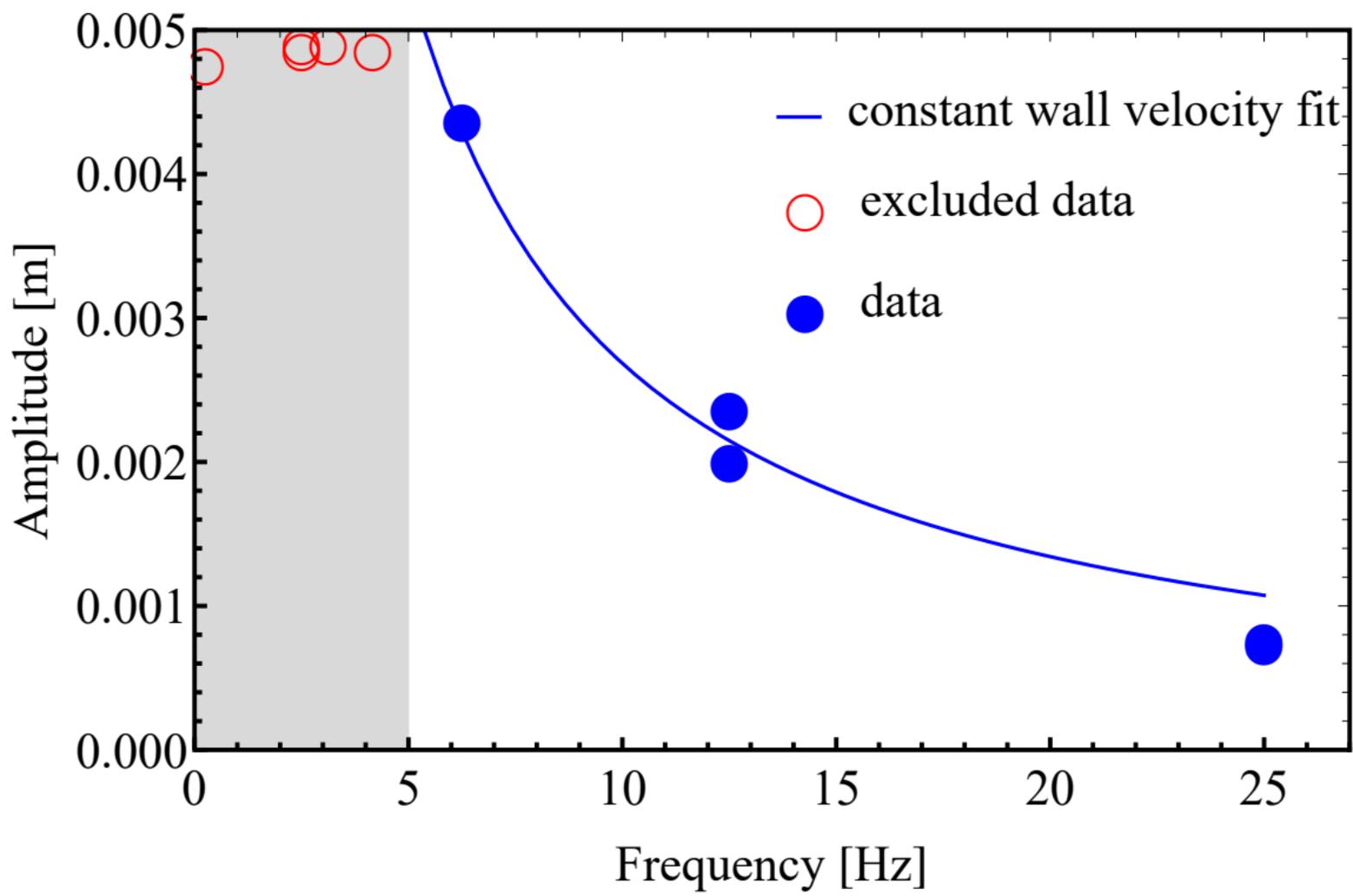

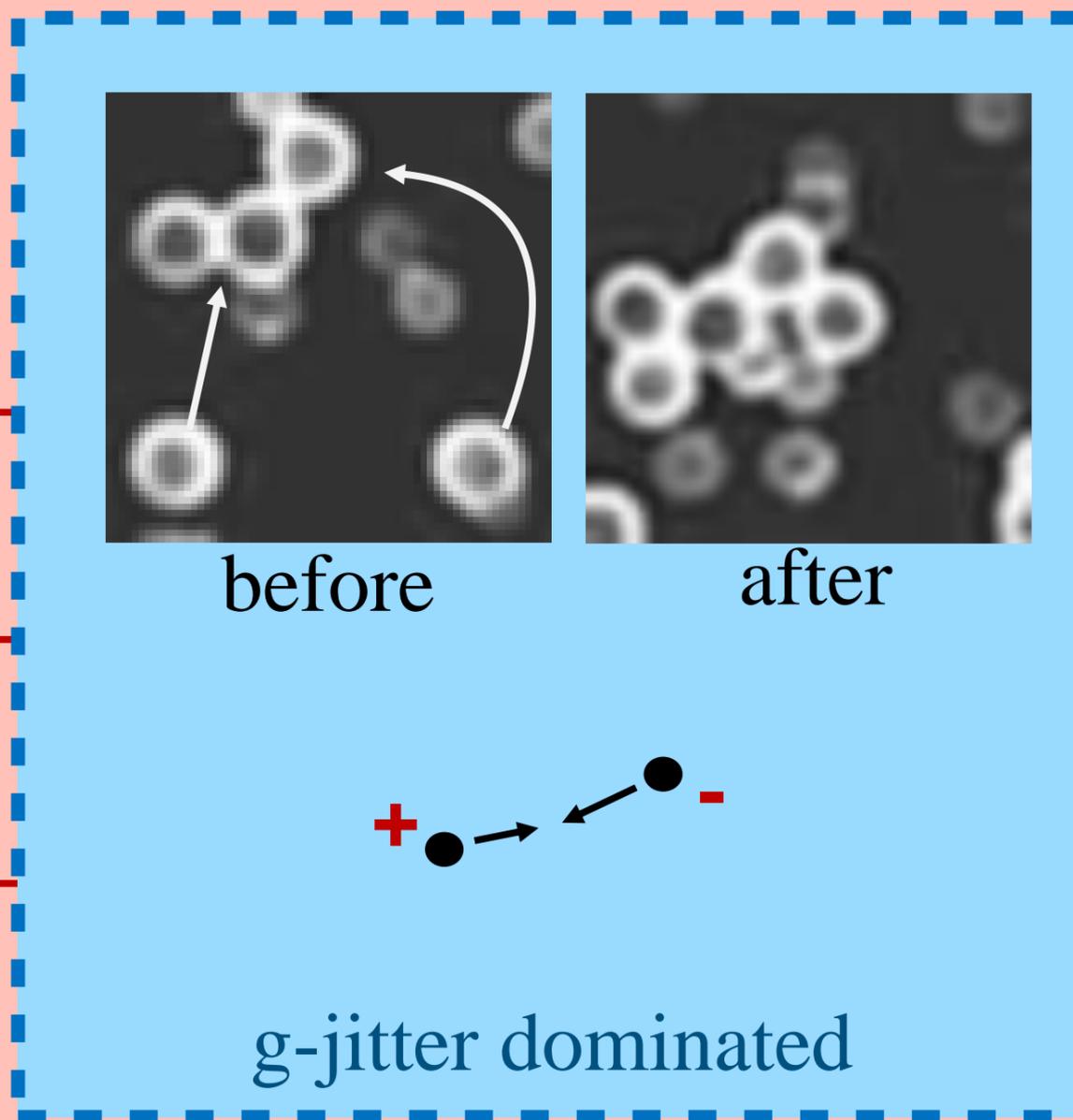

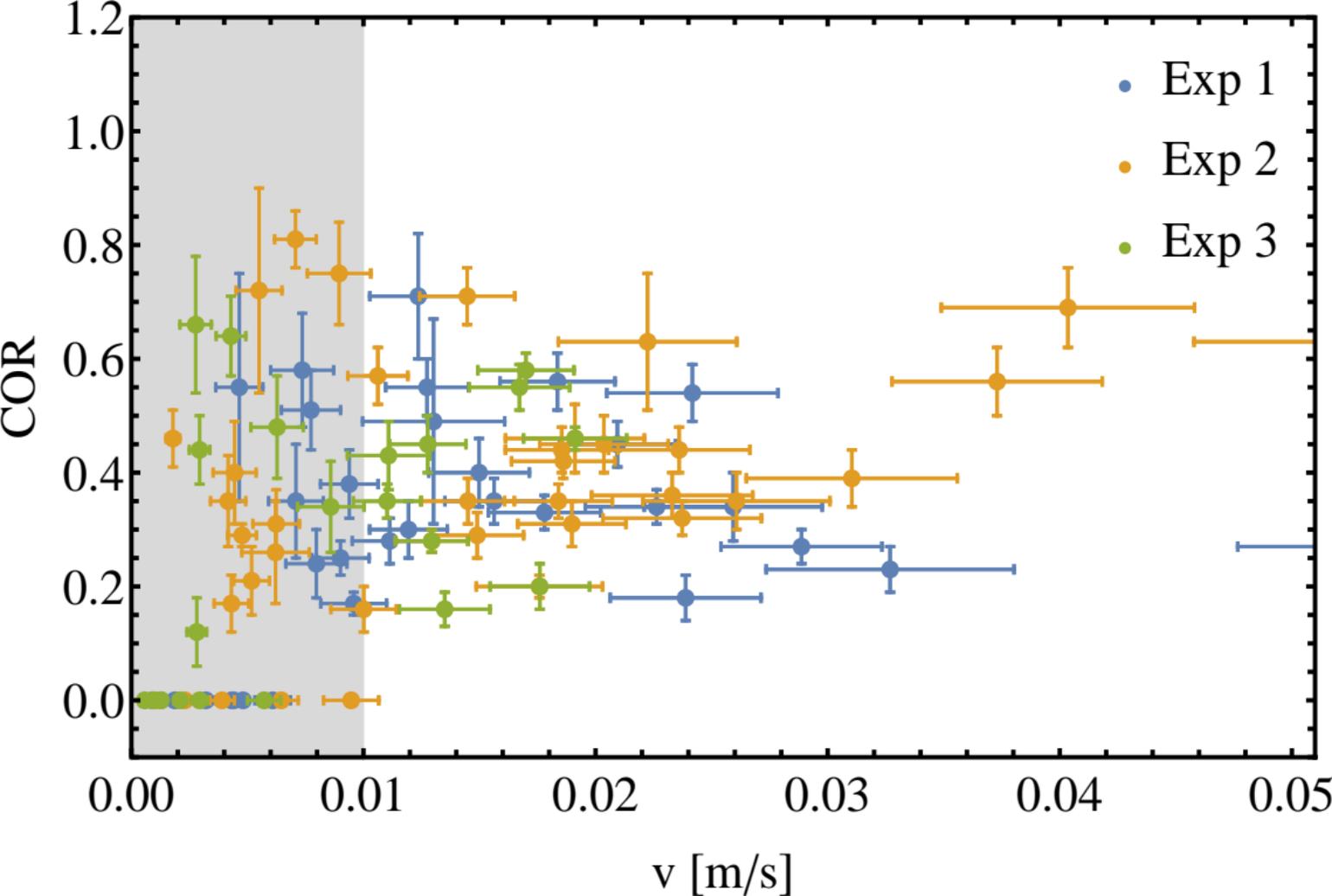

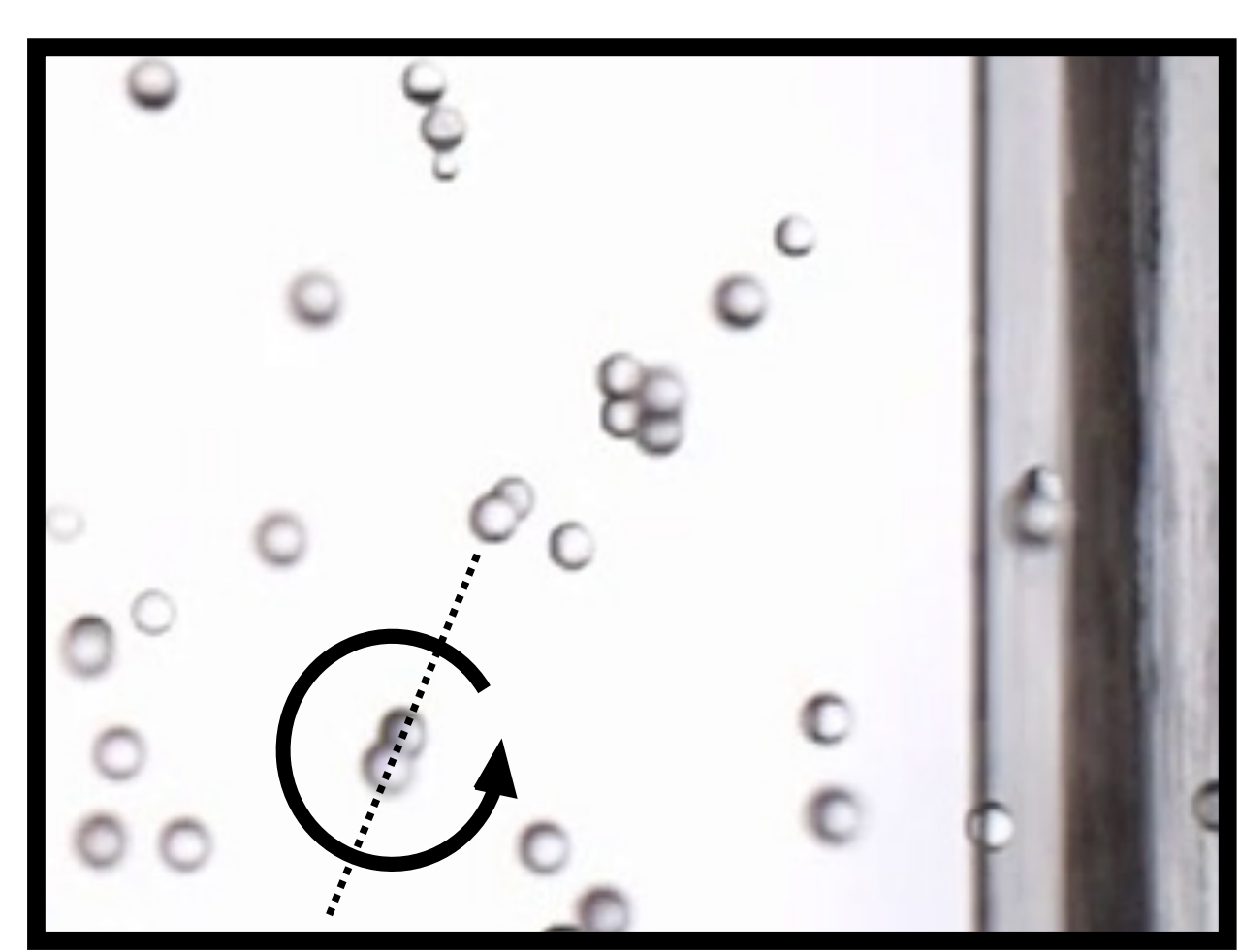 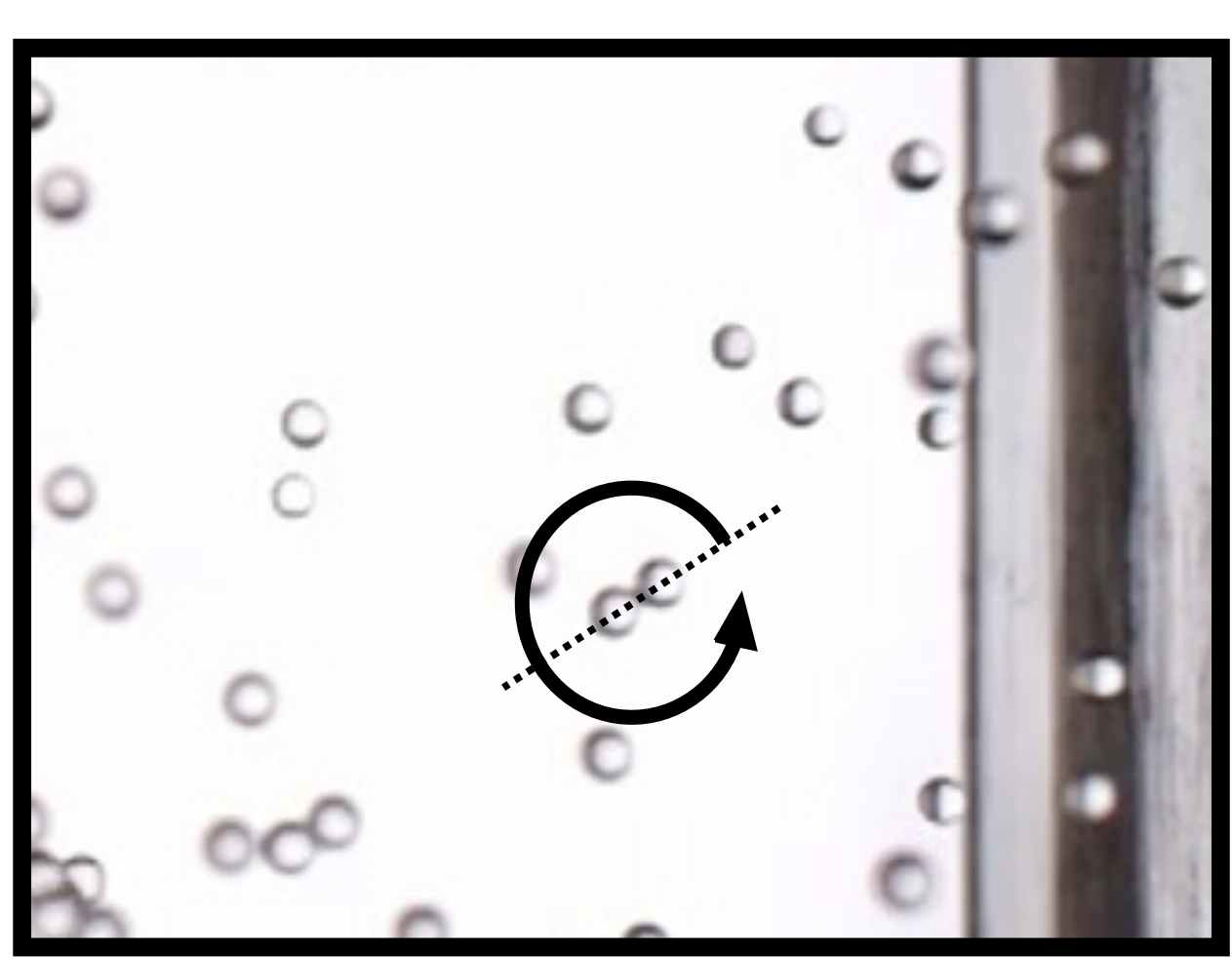 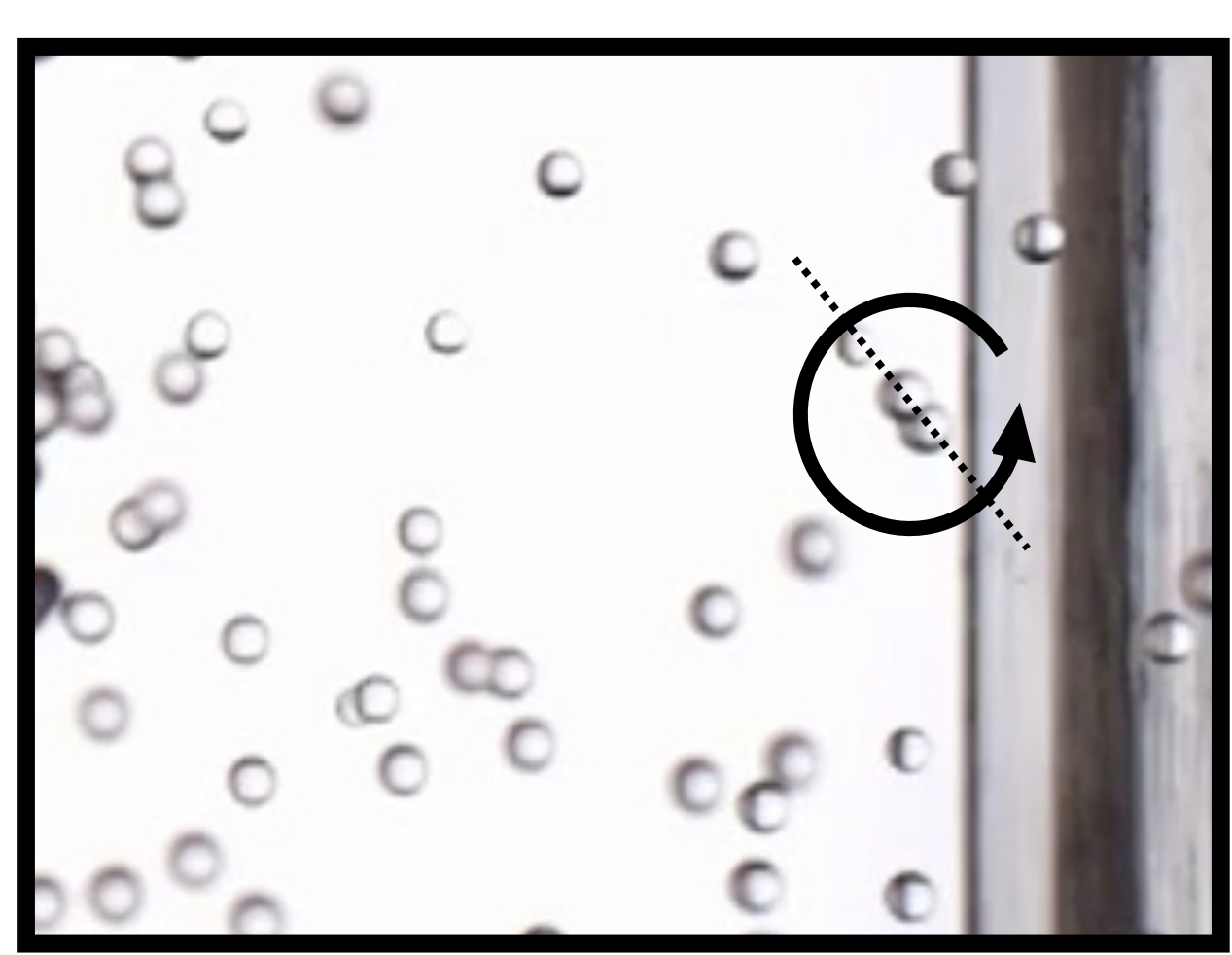

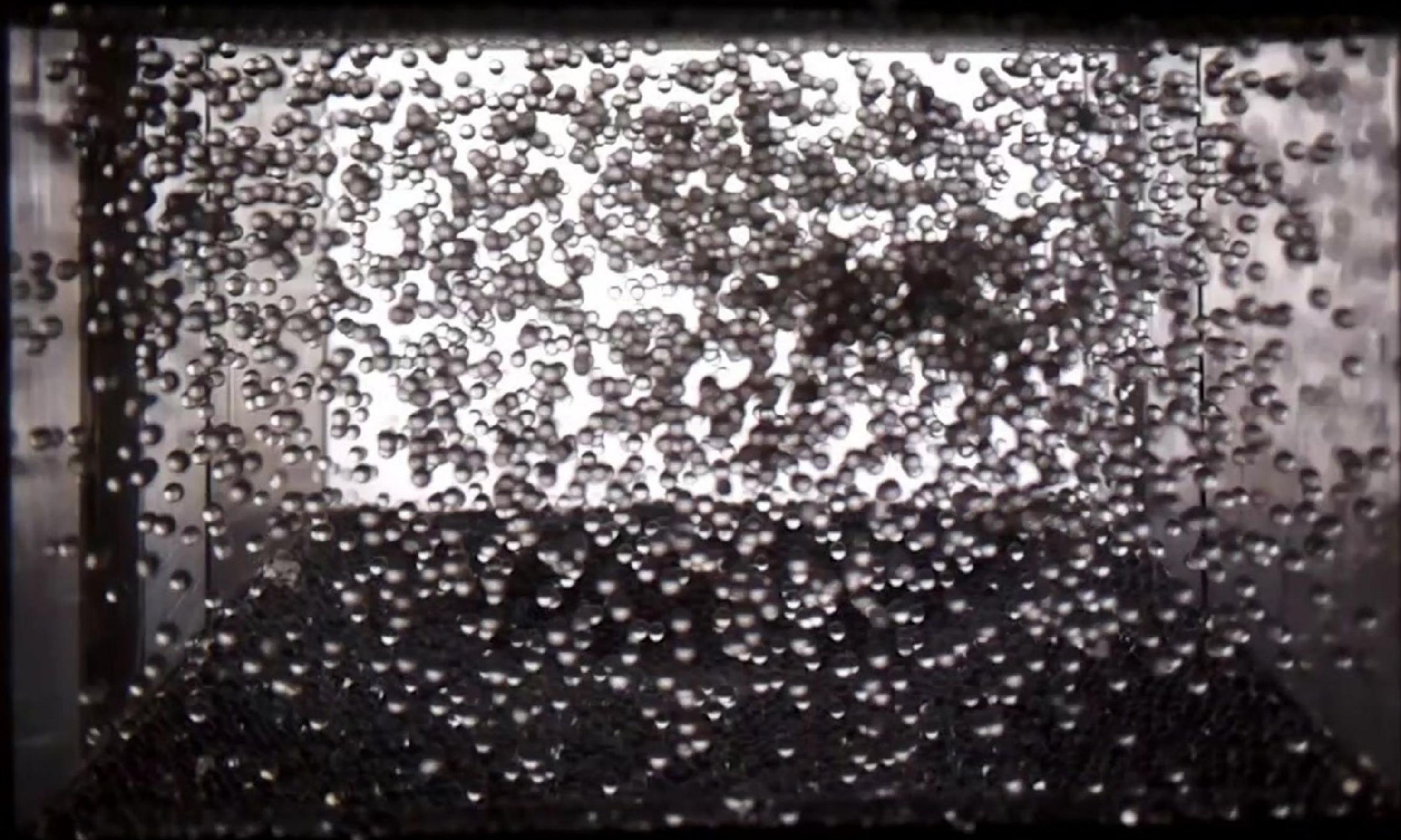

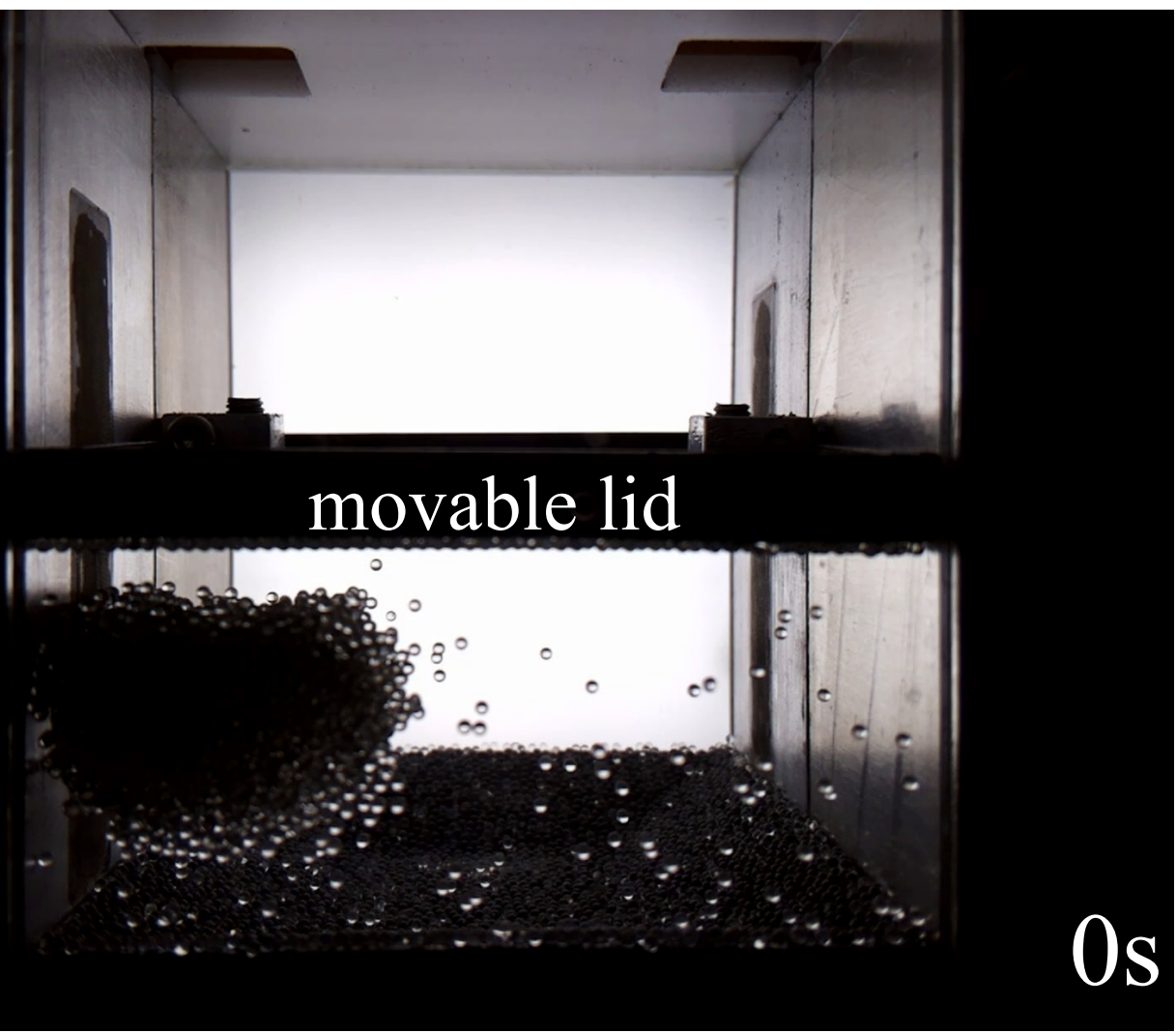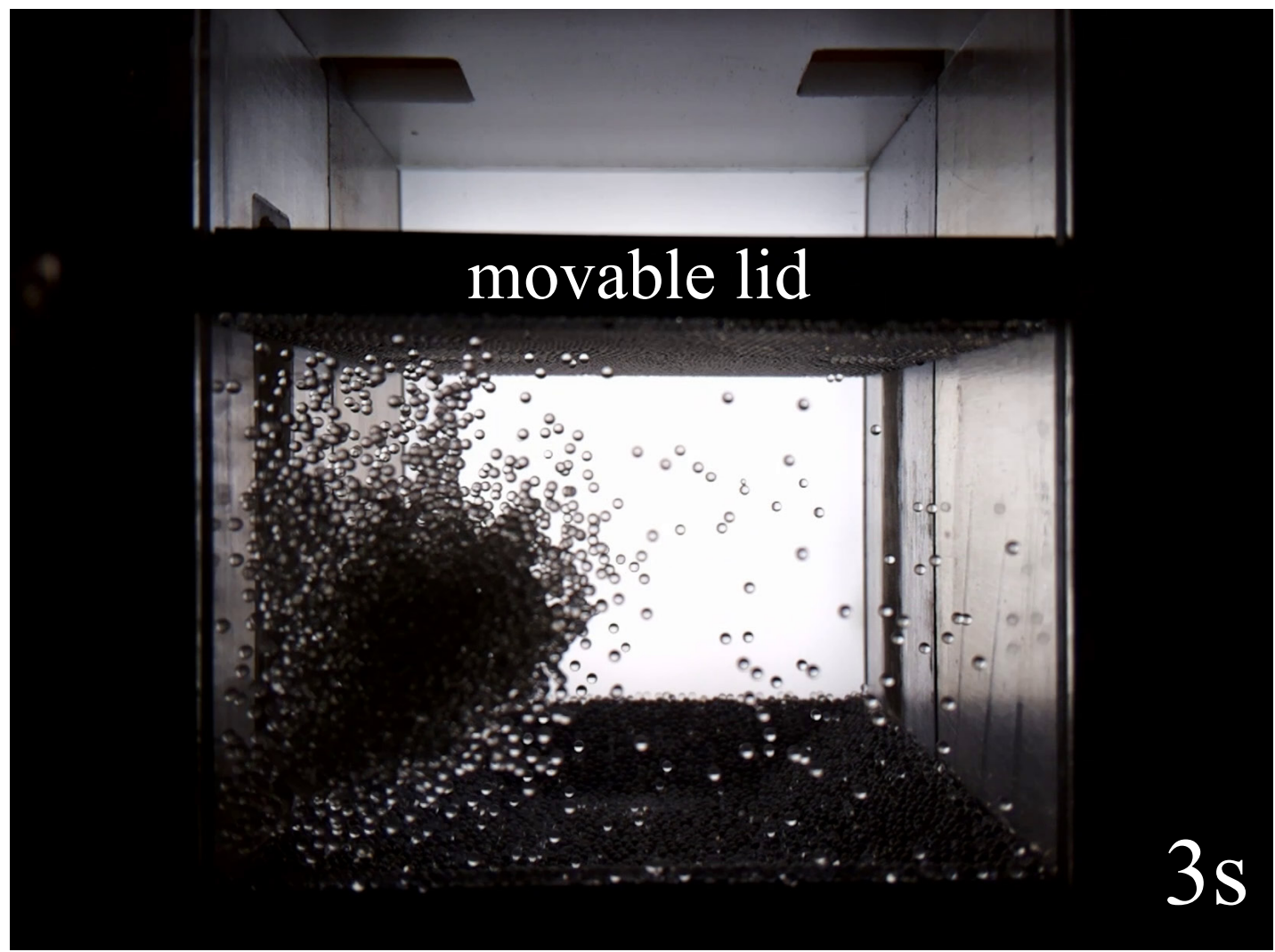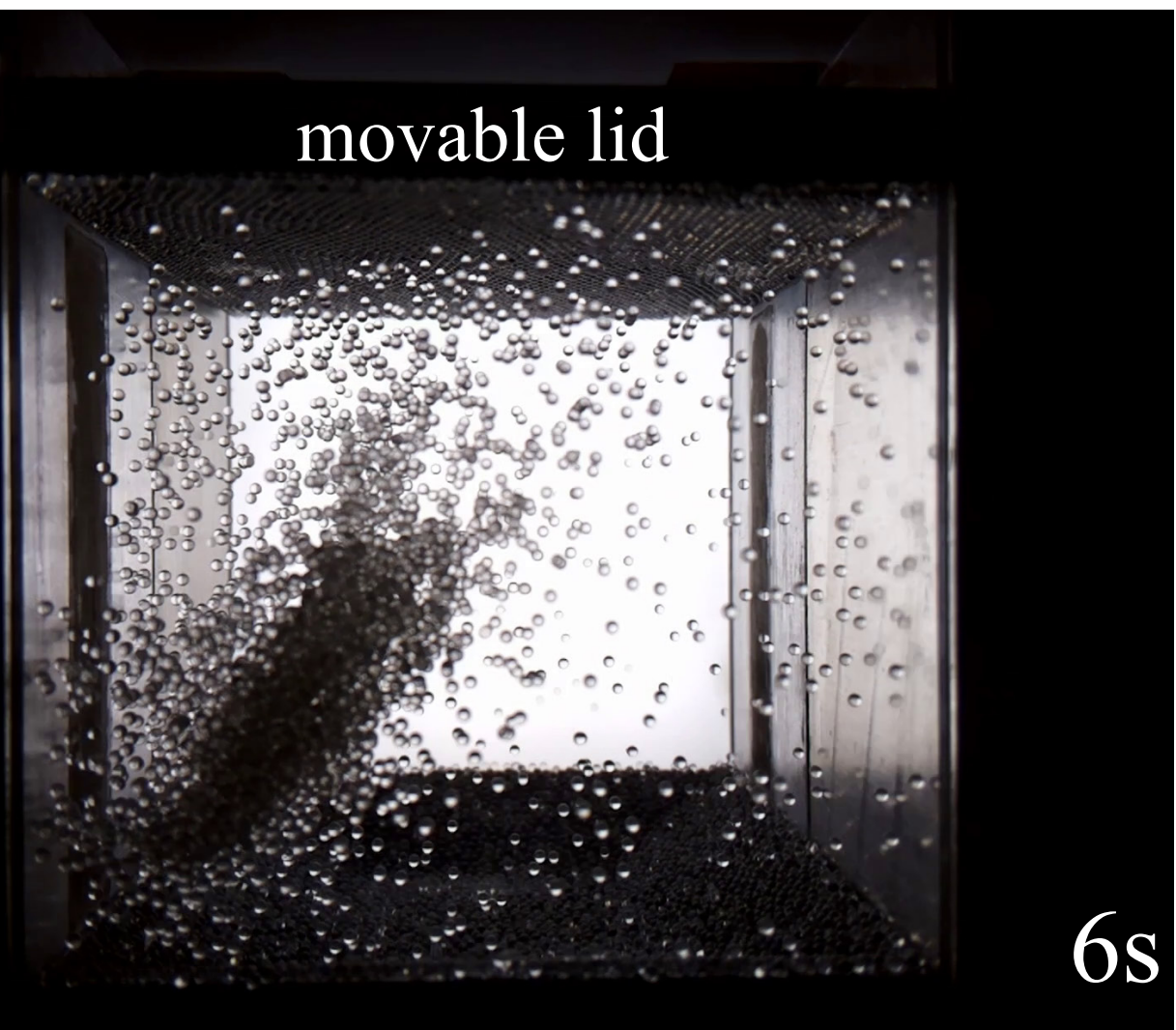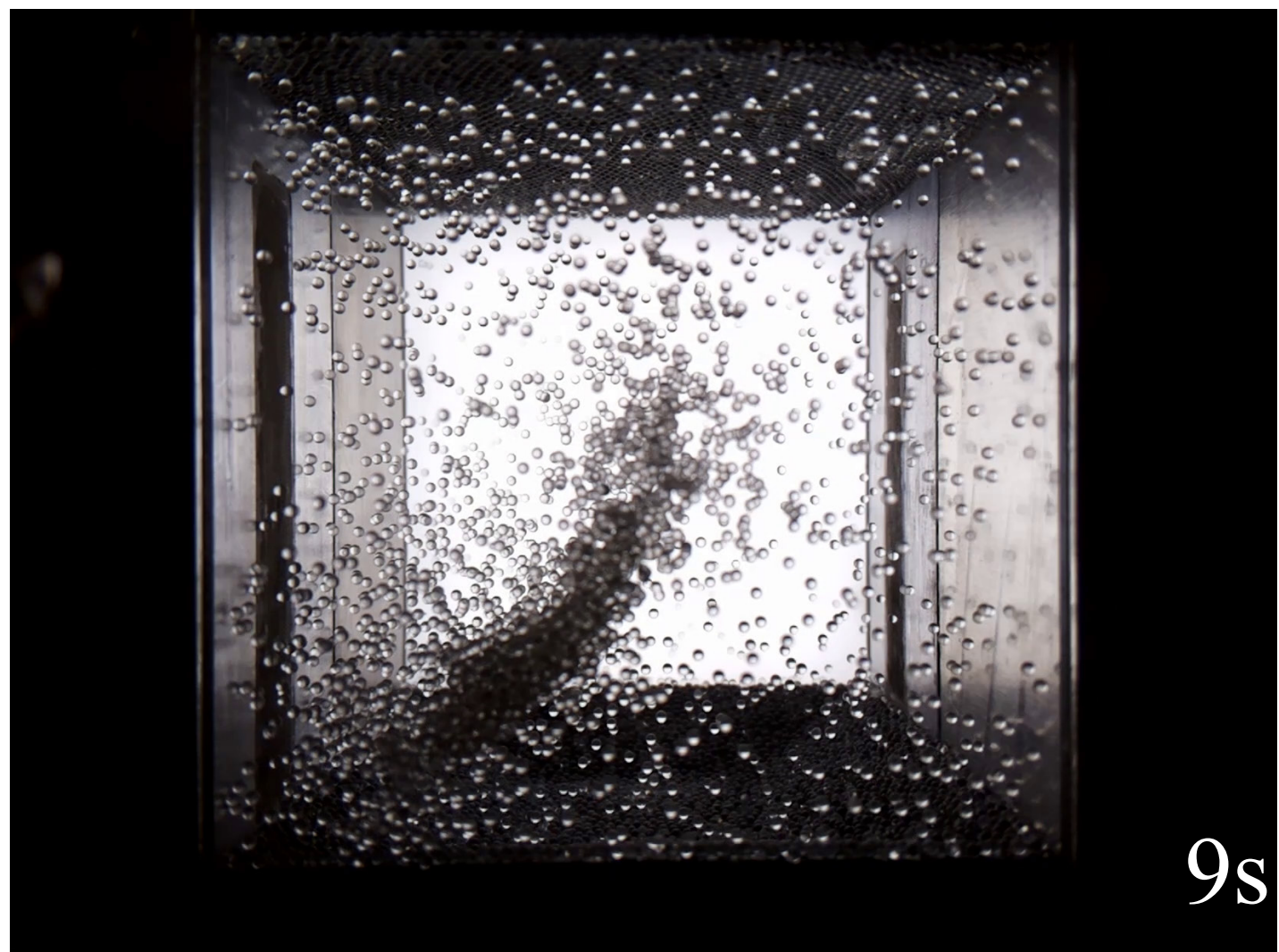

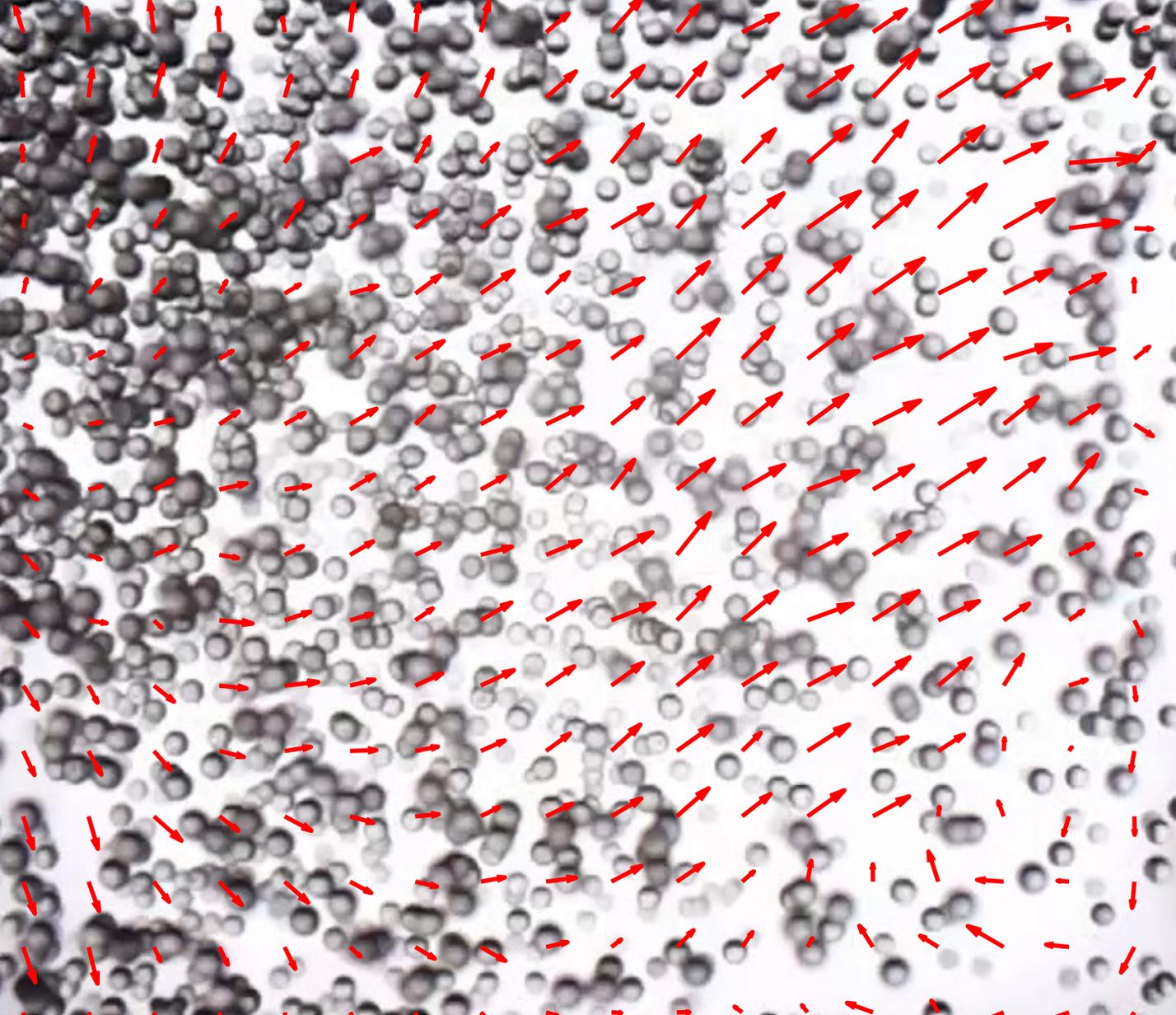

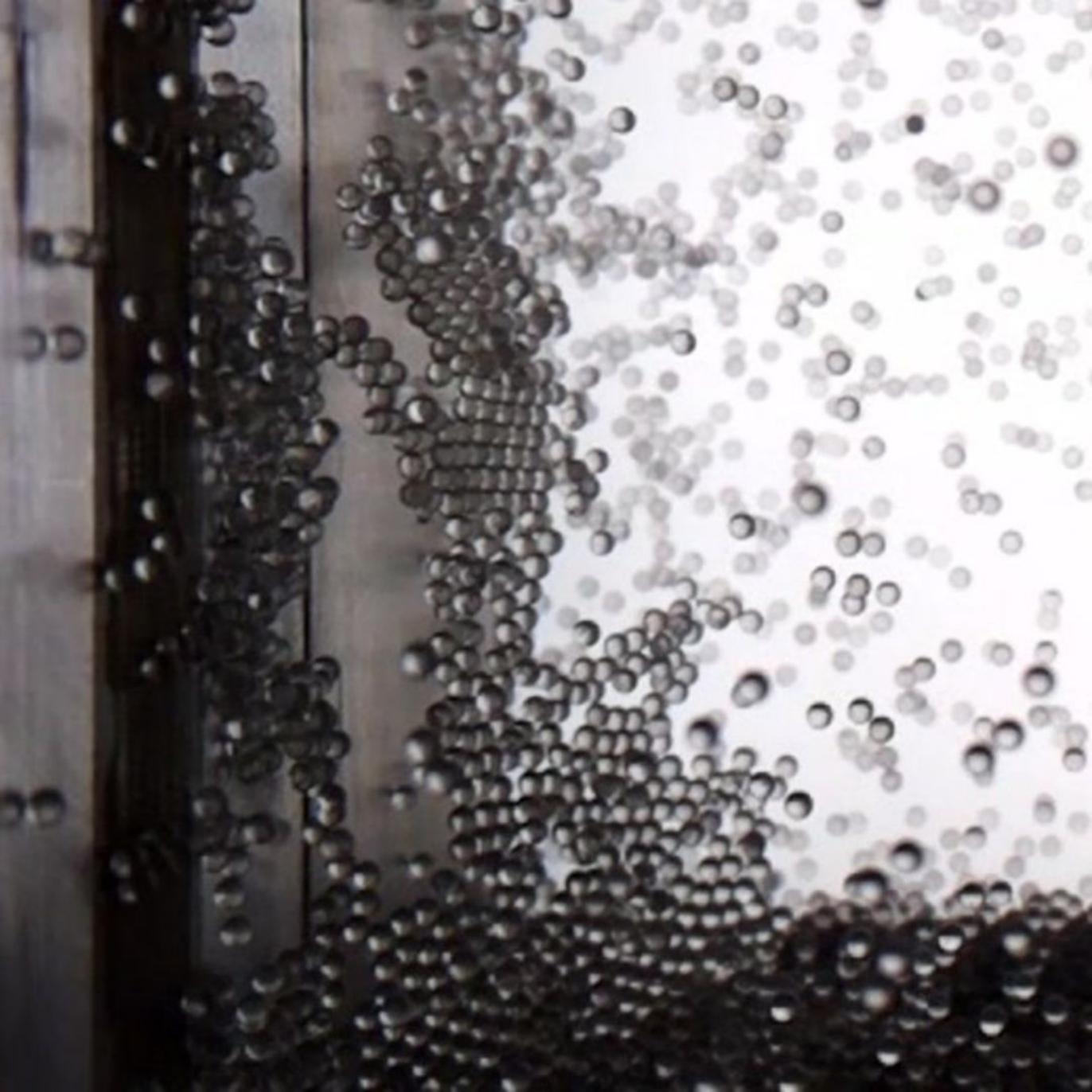